\documentclass[twocolumn,3p,number,sort&compress]{elsarticle}
\usepackage[utf8]{inputenc}

\usepackage{wrapfig, subcaption, graphicx, xcolor, placeins, siunitx, hyperref} 
\usepackage{amsmath, amssymb, bbm, bm}

\makeatletter
\def\ps@pprintTitle{%
 \let\@oddhead\@empty
 \let\@evenhead\@empty
 \def\@oddfoot{}%
 \let\@evenfoot\@oddfoot}
\makeatother

\graphicspath{{Figures/}} 

\begin{document}

\begin{frontmatter}
\title{Scalable 3D printing for topological mechanical metamaterials}

\author[add1,add2]{Achilles Bergne}
\author[add1]{Guido Baardink}
\author[add3]{Evripides G Loukaides\corref{corauth}}
\ead{E.Loukaides@bath.ac.uk}
\author[add1]{Anton Souslov\corref{corauth}}
\cortext[corauth]{Corresponding author}
\ead{A.Souslov@bath.ac.uk}
\address[add1]{
    Department of Physics, 
    University of Bath, 
    Claverton Down, 
    Bath BA2 7AY, 
    UK}
\address[add2]{
    Department of Energy Conversion and Storage,
    Technical University of Denmark, 
    Fysikvej 310, 
    2800 Kgs. Lyngby, 
    Denmark}
\address[add3]{
    Department of Mechanical Engineering, 
    University of Bath, 
    Claverton Down, 
    Bath BA2 7AY, 
    UK}

\begin{abstract}
    Mechanical metamaterials are structures designed to exhibit an exotic response, such as topological soft modes at a surface. 
    Here we explore single-material 3D prints of these topological structures by translating a ball-and-spring model into a physical prototype.
    By uniaxially compressing the 3D-printed solid having marginal rigidity, we observe that the surfaces are consistently softer than the bulk. 
    However, we also find that either of two opposite surfaces can be the softest, in contrast to the topologically robust predictions of the linear model.
    Finite-element simulations allow us to bridge this gap. 
    We explore how the printing geometry and deformation amplitude could affect surface softness. 
    For small strains, we find qualitative agreement with the ball-and-spring model but, surprisingly, nonlinear deformations can select which side is softest.
    Our work contextualizes the predictions of topological mechanics for real 3D materials and their potential for cushioning applications.
\end{abstract}

\end{frontmatter}

\begin{figure*}[t!] 
    \centering
    \includegraphics[width=\linewidth]{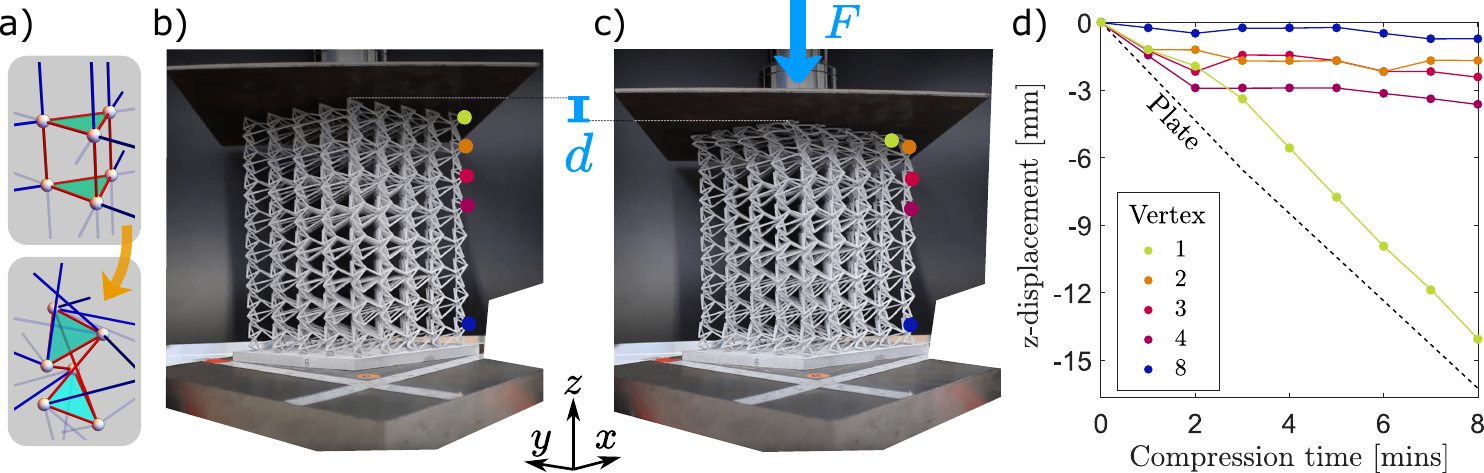}
    \caption{
    \textbf{Surface softness of the isostatic mechanical insulator.}
    (a) Design of the metamaterial, which was obtained by deforming a six-particle unit cell from the stacked kagome lattice~\cite{baardink2018localizing}. 
    (b--c) Stills from the Supplementary Video of the  $5 \times 5 \times 8$ lattice printed using Multi Jet Fusion (MJF) before (b, $t=0\ \mathrm{min}$) and after  (c, $t=8\ \mathrm{min}$) compression by a force $F$ over a distance $d$.
    Each uncompressed unit cell is $\SI{2}{\centi\metre}$ in height for a total height of $\SI{16}{\centi\metre}$. 
    The labeled points correspond to the same node across different metamaterial unit cells.
    (d) Vertical displacements of the five positions in (b--c) at 1 minute intervals (labeled in their respective colors). 
    The dashed line indicates the piston path.
    During compression, the top unit cell deforms and tracks the movement of the plate, in contrast to the small displacements of the other unit cells.
    }\label{fig:EdgeSoftness}
\end{figure*}

\section{Introduction}
Ball-and-spring lattice models are effective at describing the mechanical behavior of many naturally occurring materials. 
Here, atoms are replaced by massive point particles and their bonds are represented by harmonic springs.
More recently, these ball-and-spring lattices have been carefully crafted to exhibit exotic physical properties that only occur within these artificial structures.
A challenge is to faithfully translate the mechanical properties of these models into physical designer materials.

The proliferation of accessible Additive Manufacturing (also known as 3D printing) could offer a pragmatic solution to the challenge of fabricating lattices with many complex unit cells. 
3D printing presents many advantages. 
All available processes are inherently digital and do not require any dedicated tooling. 
Therefore, parts can be quickly produced and refined. 
For small production batches, this also translates to a low cost per item. 
In typical 3D-printing processes, little material is wasted, revealing another efficiency. 
At the same time, 3D-printing processes are capable of geometrical complexity not possible through any other manufacturing method. 
This is pertinent to 3D lattice geometries, where subtractive, transformative, and joining manufacturing methods can only be employed in the simplest cases. 

For these reasons, 3D printing is the predominant method for the production of mechanical metamaterials. Many material systems can be explored through additive manufacturing for lattices, including combining multiple materials in a single print~\cite{chen2018multi}. Similarly, 3D printing enables metallic lattices with features at different scales~\cite{zheng2016multiscale}. Through photopolymerisation of pre-ceramic monomers, high-temperature metamaterials can be created~\cite{cui_hensleigh_chen_zheng_2018}. A recent review paper shows the proliferation and the diversity of 3D printing in this domain, but also highlights that a substantial proportion of the available techniques are only available in the lab, or come at a substantial cost~\cite{askari2020additive}.  

For complex, open lattices arising from ball-and-spring models, the 3D-printing approach needs to allow for these delicate geometries.
Characteristically, theoretical ball-and-spring designs feature slender bonds connected at the nodes by perfect hinges~\cite{lubensky2015phonons,kane2014topological,baardink2018localizing,stenull2016topological}. 
For 3D printing, this idealization presents a significant fabrication challenge. 
To approximate the hinges, one option is to combine a stiff material for the springs with a second, more flexible material for the balls~\cite{bilal2017intrinsically}.
However, for intricate lattices with many unit cells, this construction (which also requires a third material for support) quickly becomes prohibitively cost and labor intensive.
Single-material 3D printing offers a scalable and affordable solution. 
Various approaches have yielded designed functionality in 3D structures despite the fabricated mechanics departing significantly from the idealized ball-and-spring models~\cite{paulose2015selective,kadic20193d,bilal2021experimental}. 
However, for single-material 3D printing to become a standard tool for realizing ball-and-spring models, further exploration of how to faithfully translate these models into a network of joined beams is required.

Mechanical metamaterials~\cite{kadic20193d,bertoldi2017flexible} form a broad class of designer matter with emergent mechanical properties. 
Typically, these materials are built from periodically repeating unit cells, which can be finely tuned to yield a vast and diverse range of macroscopic responses for numerous applications. 
Metamaterials allow for designer mechanical properties, such as elastic moduli, leading to surprising behavior including negative Poisson's ratio (or auxetic response)~\cite{burns1987negative,bertoldi2010negative,yu2018mechanical,ren2018auxetic}. Even more intricate control at the meta-atom scale leads to programmable shape change~\cite{florijn2016programmable,overvelde2016three,lei20193d}. Moreover, metamaterials can tailor dynamical behavior of vibrations~\cite{dalela2021review}, sound waves~\cite{cummer2016controlling}, and energy fluxes~\cite{wu2021brief}. More generally, metamaterials' ability to sense and respond to their environment could make them a key component in smart and animate matter~\cite{pishvar2020foundations,animate2021}.

In this work, we focus on 3D printing the specific ball-and-spring metamaterial architecture modelled in Ref.~\cite{baardink2018localizing}.
A key feature of this periodic lattice is that it acts as a 3D mechanical insulator. Specifically, the ball-and-spring material is marginally rigid in its interior (or isostatic, hosting no bulk floppy modes). 
However, its top and bottom surfaces are soft (or malleable) due to the presence of localized floppy modes.
In addition to this \emph{softness localisation}, our homogeneous lattice features a second designer property which we call \emph{surface softness asymmetry}, because the top surface is softer than the bottom. 

This surface softness asymmetry is captured by a vector $\bm{P}$, which points towards the softer face of the material.
This softness polarisation $\bm{P}$ is an example of a topological invariant, which can only take on discrete values. In general, topological mechanical metamaterials have been designed to host a variety of such invariants, defined from their bulk vibrational spectrum~\cite{susstrunk2016classification, bertoldi2017flexible,shankar2022topological}. 
The topology encodes information about the mechanics at locations where the invariant becomes ill-defined (for example at edges, boundaries between materials, and defects) according to a general principle called bulk-boundary correspondence~\cite{kane2014topological,rocklin2016mechanical}.
Due to their discrete nature, topological invariants often endow the underlying materials with robust properties.
Recently, mechanical invariants have been used to engineer scattering-free acoustic and elastic waves~\cite{yang2015topological,khanikaev2015topologically,mousavi2015topologically,nash2015topological}, as well as localized instabilities~\cite{ghatak2020observation,coulais2021topology}, failure modes~\cite{paulose2015selective,zhang2018fracturing}, and elastic deformations~\cite{kane2014topological,chen2014nonlinear,chen2016topological}.
The softness polarisation vector $\bm{P}$ that we consider endows the material with robust softness asymmetry, which is independent of the printing material and lattice dimensions~\cite{baardink2018localizing}. In this way, topological mechanical metamaterials could have wear-resistant applications in cushioning and vibration control~\cite{bilal2017intrinsically}.

In this work, we 3D print a topological mechanical insulator from a single material.
We briefly review the theory underlying these structures in Section 2.
Experimentally, we uniaxially compress our samples and observe softness localization at the material surface. 
By measuring the reaction forces at free surfaces (Section 3), we find that the softness asymmetry varies with the printing technique and the deformation amplitude. 
We then connect these experimental results to theoretical predictions using finite-element simulations (Section 4). 
In our simulations, we characterize the dependence of the reaction forces on the fabrication parameters for the 3D-printing geometry. 
We discover that the amplitude of deformation can flip the softness asymmetry, providing a link between the theoretical predictions and the experimental results.
Our work suggests a general model-to-prototype workflow for translating exotic theoretical functionality into single-material 3D prints.

\begin{figure*}[th!] 
    \centering
    \includegraphics[width=\linewidth]{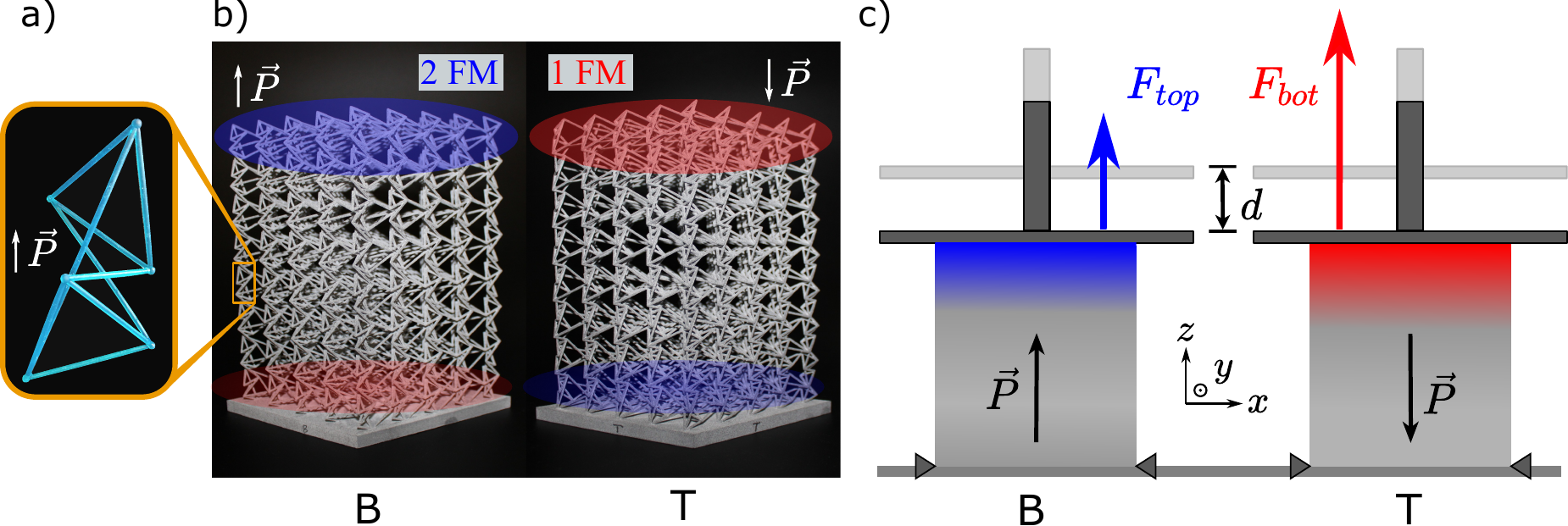}
    \caption{
    \textbf{Experimental setup.}
    (a) SLA print of a single unit cell of the structure, with polarization vector $\bm P$ indicating the top of the original structure.
    (b) Two $5 \times 5 \times 8$ Multi Jet Fusion (MJF) prints of the same structure with different orientations with respect to a rigid base plate, labeled B for base-on-bottom and T for base-on-top.
    Without the base plate, linear theory would predict two floppy modes (FM) on the top (blue) surface, and one floppy mode on the bottom (red).
    (c) The two prints behave differently under compression tests.
    Because the base plate pins the surface to restrict translations and rotations in any direction, the deformation will happen mostly at the opposing surface, which is in contact with the piston.
    The reaction force $F_{top}$ measures the resistance against deformations for the side (of the B print) with \emph{two} floppy modes.
    Analogously, $F_{bot}$  measures the resistance against deformations for the side (of the T print) with only \emph{one} floppy mode.
    }
    \label{fig:ExperimentalSetup}
\end{figure*}

\section{Topological theory of surface softness}
A mechanical insulator is a ball-and-spring lattice which does not support any floppy modes in its bulk.
By floppy modes, we mean collective ball displacements which incur no energy cost (at the lowest, quadratic order).
In other words, for sufficiently small deformations, floppy modes are the softest~\cite{lubensky2015phonons}.
In a sub-isostatic configuration, the degrees of freedom for the balls outnumber the constraints imposed by the springs. The Maxwell counting argument~\cite{maxwell1864calculation,calladine1978buckminster,lubensky2015phonons} guarantees that any such under-constrained configuration must host floppy modes.

If we cut springs from an isostatic (i.e., marginally rigid) mechanical insulator, we liberate a floppy mode for every spring that we remove.
For example, if we cut an infinite isostatic insulator down to a slab with finite height, the resulting structure features $m\times N$ floppy modes.
Here, $N$ counts the unit cells on the top surface, and $m$ counts the cut springs per unit cell, which were formerly connected to their vertical neighbors.
For an infinitely wide slab (i.e., $N\to\infty$), these floppy modes organize themselves into $m$ bands in the vibrational spectrum.
The insulating bulk guarantees that each band must be localized entirely at either the bottom or the top of the finite lattice. This argument leads to the crucial prediction that the structure is softer at the surface than in the bulk.

How can we obtain such a marginally rigid mechanical insulator? 
Although these structures are generic in 2D~\cite{rocklin2016mechanical}, they have proven elusive in 3D. Reference~\cite{baardink2018localizing} describes the design of a parametrized family of periodic isostatic lattices in 3D (based on the stacked kagome lattice, see Fig.~\ref{fig:EdgeSoftness}a) and subsequent discovery of a mechanical insulator within that parameter space. When printed and compressed, this structure presents a surface which is much softer than its rigid bulk. The differences between the uncompressed (Fig.~\ref{fig:EdgeSoftness}b) and compressed (Fig.~\ref{fig:EdgeSoftness}c) structure are highlighted in Fig.~\ref{fig:EdgeSoftness}d. 
The nodes near the top edge of the structure are displaced significantly more than the nodes within the bulk. 
This indicates the presence of a localized topological soft mode at the edge of the structure.

Since our lattice features $m=3$ vertical bonds between unit cells, the top and bottom surfaces host an odd number of bands of floppy modes.
These bands must be attached to either the top or the bottom, and therefore, one of the surfaces must necessarily be softer than the other. Reference~\cite{baardink2018localizing} computes a topological vector $\bm P$ from the geometry of the bulk.
This softness polarization vector $\bm P$ points towards the surface which we dub to be the \emph{top}. Bulk-boundary correspondence~\cite{kane2014topological,rocklin2016mechanical} then predicts that the top surface hosts more floppy modes than the bottom.

\begin{figure}[tbh!] 
    \centering 
    \includegraphics[width=78mm]{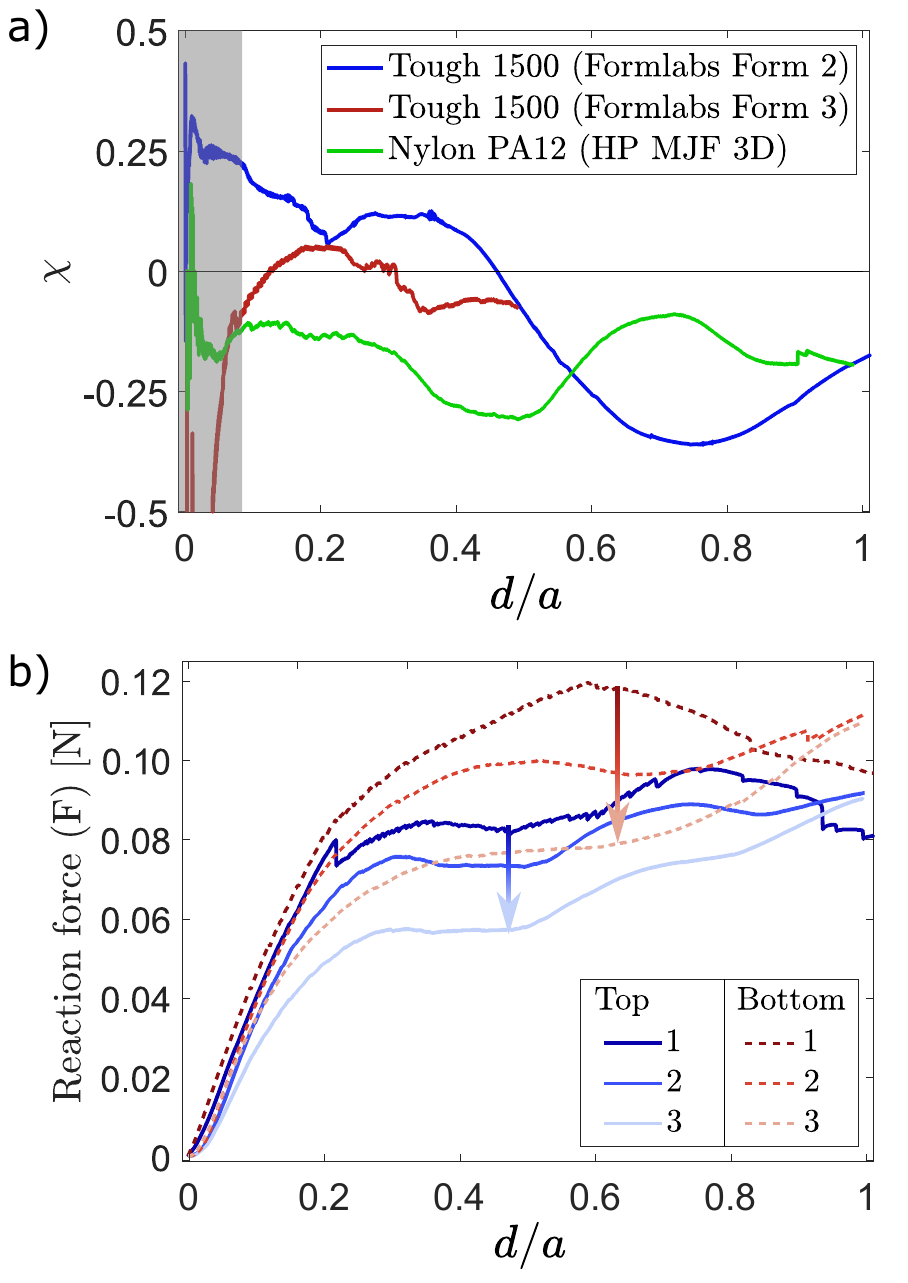}
    \caption{
    \textbf{Experimental data from the universal testing machine.}
    (a)  Softness asymmetry parameter $\chi$[$\equiv \ln(F_{bot}/F_{top})$] as a function of the deformation parameter $d/a$, where $d$ is the piston displacement and $a$ is the height of the unit cell. The three curves are labeled by the constituent resinous material (with the type of 3D printer in brackets). 
    (b) Reaction force against the deformation parameter $d/a$ for the nylon (MJF) print from part (a). The reaction forces measured at the top (solid blue line) and bottom (dashed red line) surfaces are shown for three consecutive repetitions (labeled 1, 2, and 3). 
    The downward arrows indicate that the structure softens after each compression.
    }
    \label{fig:ExperimentGraphs}
\end{figure}

\section{Softness asymmetry experiments}
To independently probe the softness on both sides of the structure, we prepare two copies, each attached to a rigid base plate. 
A base plate on the bottom (B) or top (T) side of the lattice locally restricts movement. The base plate therefore eliminates the floppy modes attached to one side and allows us to characterize the floppy modes on the other side in isolation, see Fig.~\ref{fig:ExperimentalSetup}a--b. 
We quantify the isolated softness of the top (bottom) surface by measuring the reaction force $F_{top}$ ($F_{bot}$) under compression by a universal testing machine, as a function of the piston displacement $d$, see Fig.~\ref{fig:ExperimentalSetup}c. 
To express which side is softer and by how much, we use the dimensionless softness asymmetry parameter defined as
\begin{equation}
    \chi \equiv \ln\frac{F_\textrm{bot}}{F_\textrm{top}}.
\end{equation}
The linear theory predicts that the top side will resist displacement less that the bottom, corresponding to a positive $\chi$. Conversely for a negative $\chi$, it is the bottom side which will offer less resistance.

Several samples with $5\times5\times8$ unit cells of the structure were (single-material) 3D printed. For simplicity, we replaced the springs of the ball-and-spring model by beams of a uniform thickness. We compare different prints to find how the mechanical properties change with the underlying material and printing process. 
Fig.~\ref{fig:ExperimentGraphs}a shows softness asymmetry $\chi$ for three different combinations of constituent resin material and 3D printer.
Initial prints were made on the Formlabs Form 2 and Formlabs Form 3 printers using Formlabs' Tough 1500 resin (with unit-cell height $a=\SI{10}{\milli\metre}$).
These printers use Stereolithography (SLA), whereby a UV-sensitive polymer resin is solidified by selective illumination using a UV-laser to form the individual layers of the print. After each layer is printed, the structure has to be sheared from the bottom of the plastic vat containing the resin. The shearing step avoids new layers of resin sticking to the printer but can deform slender structures when the resin is still soft. The Form 3 printer has a more delicate shearing step than the Form 2, thus reducing the risk of beams being deformed.
We compared these Formlabs prints with Nylon PA12 prints made using an HP Jet Fusion 5200 3D printer (with $a=\SI{20}{\milli\metre}$), which uses Multi Jet Fusion (MJF).
In this technique, each solid layer is created by applying fusing agents to the nylon powder and heating to create the final structure. An advantage of this technique is that the printed object is supported by the powder below and inside it, reducing the need for supports and enabling the printing of more slender structures.
As shown in Fig.~\ref{fig:ExperimentGraphs}a, we observe that the behavior of $\chi$ is dependent on the printing method used. These inherent differences could result from printing  tolerances producing lattices with slightly different mechanical properties.

Due to large deformation amplitudes, the
reaction forces show ageing of the material under repeated loads. However, the trends of the compression test remain, as shown in Fig.~\ref{fig:ExperimentGraphs}b.
At first, the reaction forces increase sharply until displacement reaches $d\approx0.2a$, above which the reaction forces vary, increasing and decreasing at different strain levels. 
Under repeated compression (labels 1 through 3) we observe a softening of the structure: the reaction force for each surface decreases.
We hypothesize that this softening is caused by several beams irreversibly buckling during the initial loading. This effect can be observed by comparing the number of deformed beams in the compressed (Fig.~\ref{fig:EdgeSoftness}c) and uncompressed (Fig.~\ref{fig:EdgeSoftness}b) structures.
Additional softening could originate from failures or micro-cracks of the beams.

The parameter $\chi$ involves differences of the (logarithms of) forces that include significant measurement error.
For small displacements, the measurements are dominated by these errors,
so the most reliable results are for displacements $d/a$ above $0.1$ (excluding the gray area in Fig.~\ref{fig:ExperimentGraphs}a).
The region where the linear theory applies is below this small displacement threshold.
To bridge this gap between theory and experiment, we turn to finite-element simulations.

\begin{figure}[t!] 
    \centering
    \includegraphics[width=0.9\columnwidth]{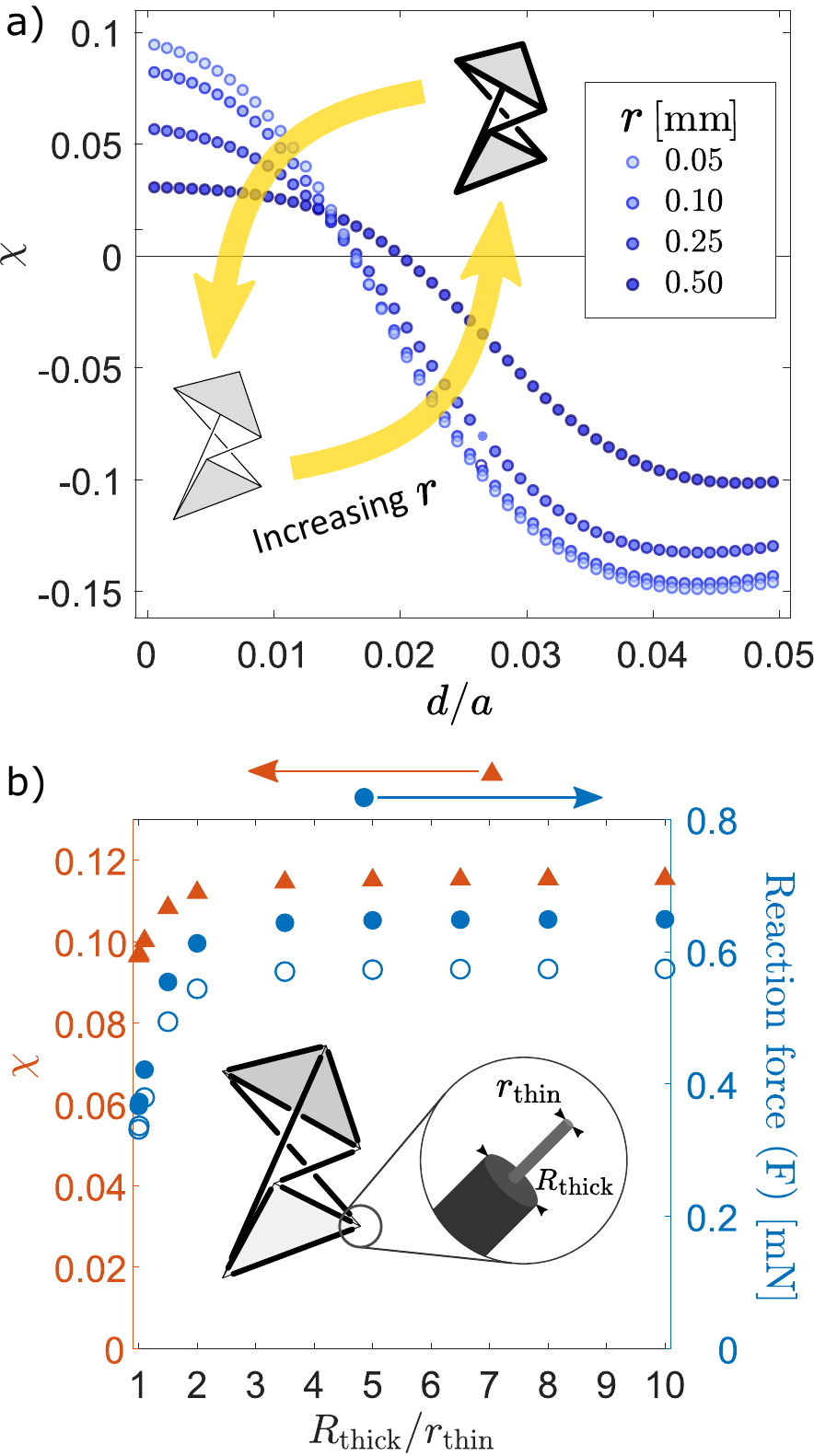}
    \caption{
    \textbf{Simulations of lattices with various beam geometries.}
    (a) Simulation results for softness asymmetry $\chi$ for a lattice with $5 \times 5 \times 8$ unit cells. We vary the beam thickness $r$ from \SI{0.05}{\milli\metre} to \SI{0.5}{\milli\metre}, while keeping the thickness uniform along the beam. 
    (b) Softness asymmetry (plotted in triangles against left vertical axis) as well as reaction forces for both bottom (plotted in filled circles against right vertical axis) and top (plotted in open circles against right vertical axis) surfaces for a $5 \times 5 \times 5$ lattice with $d=0.01a$. Here the beams have a smaller thickness $r_\textrm{thin}$ (fixed at $\SI{0.1}{\milli\metre}$) near the joints and a larger thickness $R_\textrm{thick}$ in the middle, which we vary.
    }
    \label{fig:BeamDependence}
\end{figure}

\begin{figure*}[t!] 
    \centering
    \includegraphics[width=16cm]{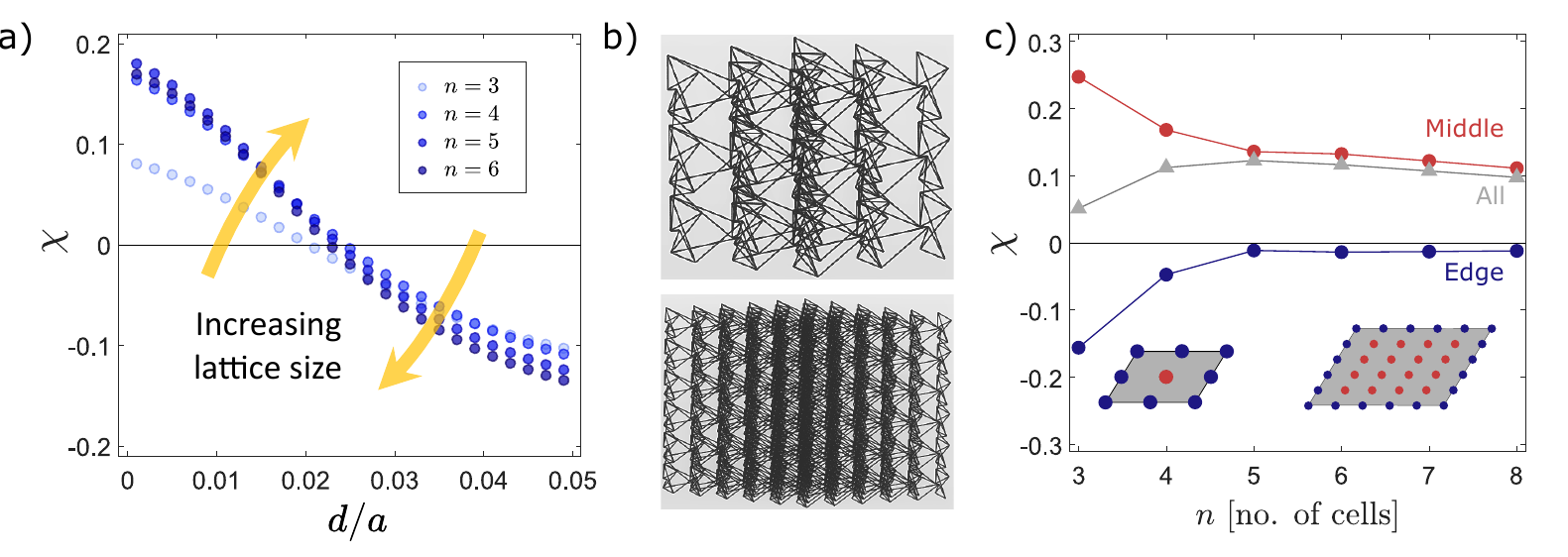}
    \caption{
    \textbf{Simulations of lattices for various sizes.}
    (a) Softness asymmetry $\chi$ as a function of deformation parameter $d/a$ for lattices containing $n\times n\times n$ unit cells. 
    (b) Schematics of the simulated beam elements for lattices with $n=3$ (top) and $n=6$ (bottom). 
    (c) Distinguishing the contributions to the total softness asymmetry $\chi$ from nodes at the edge (plotted in blue disks) and in the middle (plotted in red disks) of a surface. The total $\chi$ is plotted in gray triangles, for $d=0.01a$, $R_\textrm{thick}= \SI{0.5}{\milli\metre}$, $r_\textrm{thin} = \SI{0.1}{\milli\metre}$, and $E = \SI{200}{\mega\pascal}$. 
    The inset shows schematic representations of a $3\times3$ face and a $6\times6$ face with corresponding colors for nodes at the edges and in the middle.
    }
    \label{fig:LatticeDependence}
\end{figure*}

\section{Softness asymmetry simulations}
We simulate the structure using the Finite Elements Analysis software Abaqus. 
To efficiently simulate large lattice sizes, the beams are modeled using beam elements rather than solid elements. The $z$-dimension of each unit cell is set to $\SI{20}{\milli\metre}$ and each beam segment is $\SI{0.3}{\milli\metre}$ long.
To obtain results comparable to experiments with the universal testing machine and base-plate setup, we implement asymmetric boundary conditions. 
The nodes of the face corresponding to the base plate have all translations and rotations frozen. By contrast, the nodes on the free surface have free translations in the $x-y$ plane and free rotations, but the translation in the $z$-direction is constrained to be the displacement $d$.

Fig.~\ref{fig:BeamDependence}a plots $\chi$ for the $5\times5\times8$ lattice with varying beam thickness $r$.
For small displacements, the positive values of the asymmetry $\chi$ correspond to a top surface which is softer than the bottom, in agreement with linear theory. 
However, for displacements greater than $d/a \approx 0.01$, the sign of $\chi$ flips.

We observe that thicker beams correspond to smaller values for the magnitude $|\chi|$ of the softness asymmetry.
However, the displacement $d^*$ at which $\chi$ switches sign remains approximately the same.
Changing the Young's modulus between $1.5$ and $\SI{2.7}{\giga\pascal}$ or changing the overall length scale $a$ does not effect $\chi$, as shown in the Supplementary Material.
This demonstrates how softness asymmetry can be selected independently of the underlying materials properties, and therefore independently of parameters such as heat or electrical conductivity.  

The original ball-and-spring model assumes vanishing bending rigidity of the springs, so that the nodes act as perfect hinges.
When printing using only a single material, one could use tapered or stepped beams, which are thicker in their center than at their endpoints. In simulations, we test these design choices by splitting the beams into three sections: the thick middle section consisting of $80\%$ of the beam length (thickness $R_\textrm{thick}$), and the two thin end sections each consisting of $10\%$ of the beam length (thickness $r_\textrm{thin}$). This creates a stepped thickness profile to approximate perfect hinges, which we quantify using the parameter of thickness ratio $R_\textrm{thick}/r_\textrm{thin}$.

In Fig.~\ref{fig:BeamDependence}b, we plot simulation results for both the reaction forces $F$ and the asymmetry $\chi$ as a function of this thickness ratio. We use a constant value $r_\textrm{thin}=\SI{0.1}{\milli\metre}$ of the thickness at the ends and vary the thickness ratio between 1 and 10 (for system size $5\times5\times5$ unit cells).
When the thickness ratio is largest, the beam geometry approximates a hinge and the system exhibits the largest values of the softness asymmetry $\chi$.
Surprisingly, even for uniform beams with  $R_\textrm{thick}/r_\textrm{thin} = 1$, the value of $\chi$ does not significantly decrease away from the maximum.
For thin beam ends, the attachment points are prime locations for structural failure, and these simulation results justify our 3D-printed geometries in which we use beams of uniformly thickness.

Figure~\ref{fig:LatticeDependence}a shows that $\chi$ is positive in the low strain regime for very small lattices with $n\times n\times n$ unit cells, where we take $n$ to be between $n=3$ and $n=6$.
This is surprising, because the theoretical argument~\cite{baardink2018localizing} assumes that the lattices have both a large overall height and a large transverse cross-section (i.e., $n \rightarrow \infty$). 
In the case of small $n$, the distinction between the surface and the bulk of the lattices becomes less clear, but the quantity $\chi$ remains well defined.
We also note that the displacement $d^*$ at which the asymmetry $\chi$ changes sign does not significantly vary with lattice size. 
From Fig.~\ref{fig:LatticeDependence}a, we conclude that $n = 6$ approximates well the theoretical limit of large lattice sizes.

For the smallest lattice size of $n = 3$, we do find deviations in the behavior of the asymmetry $\chi$. 
The different contributions from edge and middle nodes as $n$ is varied are plotted in Fig.~\ref{fig:LatticeDependence}c.
We hypothesize that for $n = 3$, finite-size effects dominate, because the $8$ nodes at the edge of the top face (blue in Fig.~\ref{fig:LatticeDependence}c) vastly outnumber the one node in the middle (red in Fig.~\ref{fig:LatticeDependence}c).
 The figure shows that the contributions from the middle nodes to $\chi$ are positive and the contributions of the edge nodes are negative.
Nevertheless, this small-size effect is only relevant for $n \leq 4$ and the the asymmetry $\chi$ is determined by the middle nodes for larger lattice sizes.

\section{Concluding remarks}
In summary, we 3D printed a topological mechanical insulator and experimentally observed  softness localization at its surfaces.
Within compression experiments, the softness asymmetry varies between printing methods, in contrast to the material independence predicted by the linear theory.
In order to bridge this gap between theory and experiment, we performed finite-element simulations.
Numerically, for small strains we confirm that the softness asymmetry parameter $\chi$ is material independent and positive.
Surprisingly, we find that $\chi$ consistently changes sign beyond a critical strain, providing a mechanism for the observed variations across experiments.

In general, designer ball-and-spring geometries can feature functionalities which are challenging to carry over into 3D prints.
For example in our geometry, the constituent beams must be slender, but all 3D-printing techniques struggle with excessive slenderness.
Increasing thickness would lead to collisions between the beams, presenting a hard design constraint.
In addition, a support structure would be practically impossible to remove for such a dense lattice.
Finally, our geometry has an extreme diversity of beam orientations,
so that reorienting the structure to make every thin beam nearly vertical is infeasible.
These fabrication issues have led to deviations of the physical structures from the design, including a few broken beams. 
In addition, in the Stereolithography (Formlabs) samples, the force involved in shearing the printed layer from the resin tank also resulted in a slight overall tilt. 
This tilt is difficult to predict and depends on the placement and orientation of the sample relative to the build plate. 
Despite these challenges, our samples have allowed for an experimental quantification of the reaction forces and their decrease due to damage from repeated compressions.

Our experimental results allow us to consider broad design principles for marginally rigid, topological lattices.
Marginal rigidity leads to a material which is stiffer in the bulk than on either surface. 
Under uniform compression, the linear regime predicts one side to compress more than the other. 
However, at higher strains, the sign ambiguity of the asymmetry $\chi$ allows either side to experience most of the deformation.
We therefore predict these materials to display complex collapse behavior.
Such homogeneous, light, and wear-resistant mechanical metamaterials pave the way for practical applications from cushioning to mechanical damping.

\section*{Declaration of competing interest}
The authors declare that they have no known competing financial interests or personal relationships that could have appeared to influence the work reported in this paper.

\section*{Acknowledgments}
A.S.~acknowledges the support of the Engineering and Physical Sciences Research Council (EPSRC) through New Investigator Award No.~EP/T000961/1 and of the Royal Society under grant No.~RGS/R2/202135.

\clearpage\noindent
\textbf{\large Supplementary Material:\\[\jot] 
Scalable 3D printing for topological mechanical metamaterials}

\setcounter{equation}{0}
\setcounter{section}{0}
\makeatletter
\renewcommand{\theequation}{S\arabic{equation}}
\renewcommand{\thefigure}{S\arabic{figure}}
\renewcommand{\bibnumfmt}[1]{[S#1]}
\renewcommand{\citenumfont}[1]{S#1}
\setcounter{figure}{0}
\vspace{2pt}
Figure~\ref{supFig:Forces} shows the simulation results for the reaction forces $F$ for small lattices ($n=3$ and $n=6$)  with $n\times n\times n$ unit cells. 
This data is used to calculate the asymmetry $\chi=\ln(F_{bot}/F_{top})$ in Fig.~5. Both the top and the bottom surfaces for the $n=3$ lattice are softer (i.e., have smaller reaction forces) than for the $n=6$ lattice. 
At small strain, the softer surface is at the top for both lattices ($F_\textrm{bot}>F_\textrm{top}$ and, therefore, $\chi>0$). For higher strain, the bottom surface becomes the softer one ($F_\textrm{bot}<F_\textrm{top}$ and, therefore, $\chi<0$). 
In the main text, we define the displacement $d^*$ to be where $\chi$ changes sign, which is where the curves in Fig.~\ref{supFig:Forces} cross.

\begin{figure}[b]
    \centering
    \includegraphics[width=.87\columnwidth]{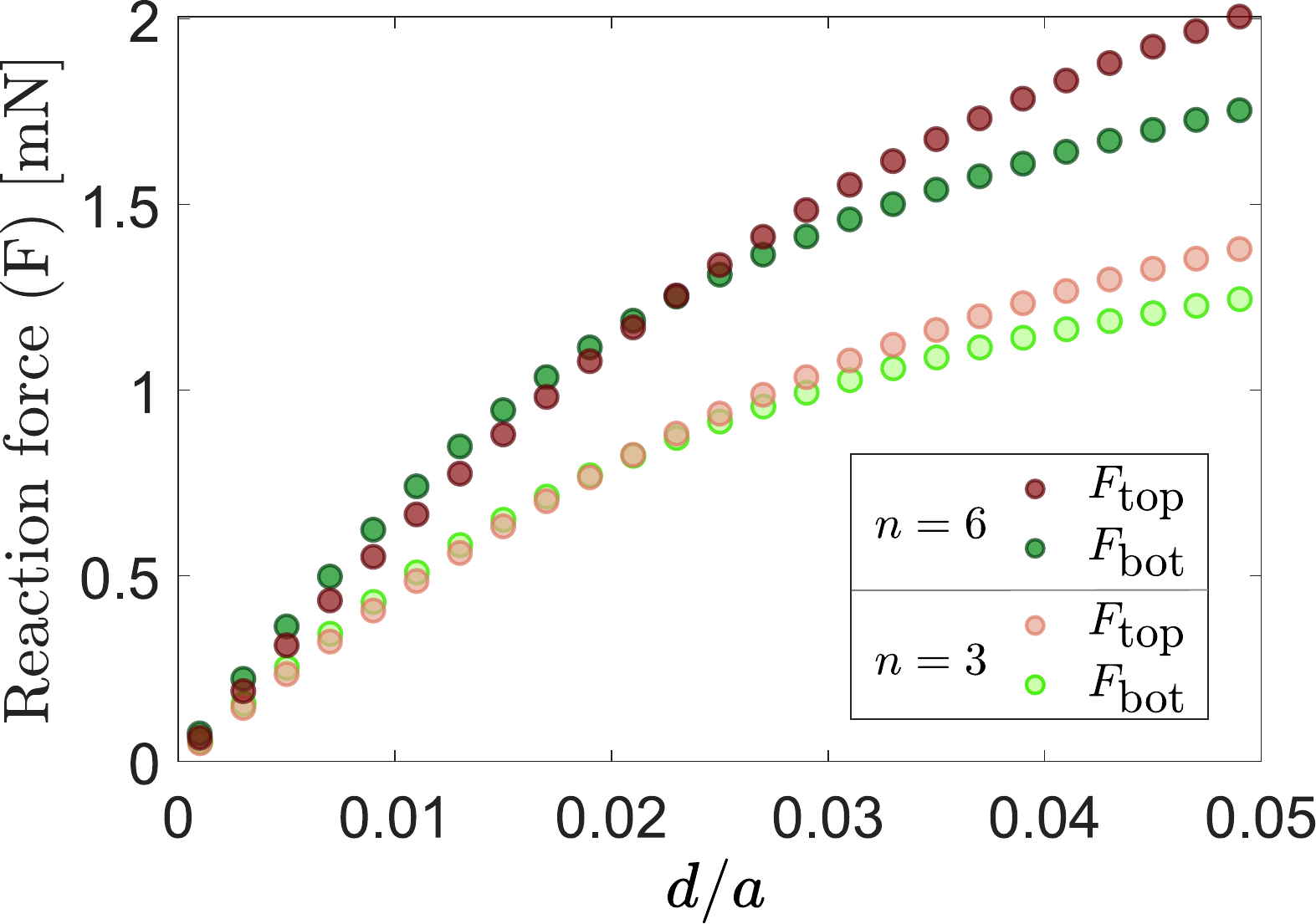}
    \caption{
    The reaction forces $F$ for $n=3$ and $n=6$, same data as Fig.~5a, plotted against deformation parameter $d/a$.
    }
    \label{supFig:Forces}
\end{figure}

\begin{figure*}[t]
    \centering
    \includegraphics[width=.9\linewidth]{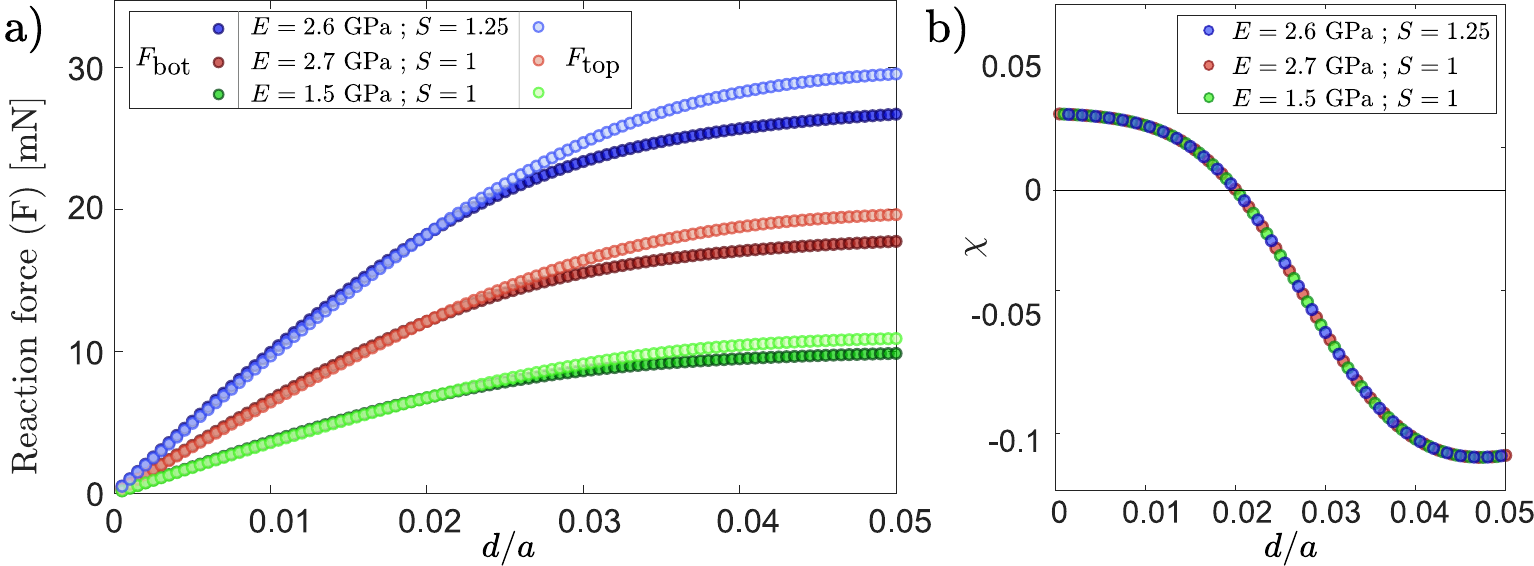}
    \caption{
    (a) Reaction forces $F$ on the bottom (dark circles) and top (light circles) surfaces as a function of the deformation parameter $d/a$ for $5 \times 5 \times 8$ lattices. 
    The colors correspond to various values of the Young's modulus, taken from Formlabs' resins: Grey Pro (blue), Tough V5 (red), and Tough 1500 resin (green). 
    In addition, for the blue data, the lattice size is scaled by the factor $S = 1.25$. The corresponding unit cell height is $a=S\times\SI{10}{\milli\metre}$ and the bond thickness is $r=S\times\SI{0.5}{\milli\metre}$.
    (b) Softness asymmetry parameter $\chi$ computed using the data in (a), plotted using the corresponding colors. All of the data points follow the same curve.
    }
    \label{supFig:Materials}
\end{figure*}

Figure~\ref{supFig:Materials} shows the simulation results for varying both the Young's modulus $E$ and the overall length scale set by the lattice spacing $a$.
The colors correspond to various values of the Young's modulus, taken from Formlabs' resins: Grey Pro (blue), Tough V5 (red), and Tough 1500 resin (green). 
We see in Fig.~\ref{supFig:Materials}a that the reaction forces depend on these materials properties.
Proportionally increasing all spatial dimensions (including the bond thicknesses), or printing in a resin with a higher Young's modulus, both result in a structure which is stiffer by an overall scaling.
This scaling cancels in the definition of the softness asymmetry parameter $\chi$, which only depends on a ration of forces.
Hence, $\chi$ does not show any differences for the three simulated lattices, as seen in Fig.~\ref{supFig:Materials}b. 
This highlights the fact that the properties of the lattice are linked to its structure, and are independent of the resins' material properties.

\begin{thebibliography}{10}

\bibitem{baardink2018localizing}
Guido Baardink, Anton Souslov, Jayson Paulose, and Vincenzo Vitelli.
\newblock Localizing softness and stress along loops in 3d topological
  metamaterials.
\newblock {\em Proceedings of the National Academy of Sciences},
  115(3):489--494, 2018.

\bibitem{chen2018multi}
Da~Chen and Xiaoyu Zheng.
\newblock Multi-material additive manufacturing of metamaterials with giant,
  tailorable negative poisson’s ratios.
\newblock {\em Scientific Reports}, 8(1):1--8, 2018.

\bibitem{zheng2016multiscale}
Xiaoyu Zheng, William Smith, Julie Jackson, Bryan Moran, Huachen Cui, Da~Chen,
  Jianchao Ye, Nicholas Fang, Nicholas Rodriguez, Todd Weisgraber, et~al.
\newblock Multiscale metallic metamaterials.
\newblock {\em Nature Materials}, 15(10):1100--1106, 2016.

\bibitem{cui_hensleigh_chen_zheng_2018}
Huachen Cui, Ryan Hensleigh, Hongshun Chen, and Xiaoyu Zheng.
\newblock Additive manufacturing and size-dependent mechanical properties of
  three-dimensional microarchitected, high-temperature ceramic metamaterials.
\newblock {\em Journal of Materials Research}, 33(3):360–371, 2018.

\bibitem{askari2020additive}
Meisam Askari, David~A Hutchins, Peter~J Thomas, Lorenzo Astolfi, Richard~L
  Watson, Meisam Abdi, Marco Ricci, Stefano Laureti, Luzhen Nie, Steven Freear,
  et~al.
\newblock Additive manufacturing of metamaterials: A review.
\newblock {\em Additive Manufacturing}, 36:101562, 2020.

\bibitem{lubensky2015phonons}
TC~Lubensky, CL~Kane, Xiaoming Mao, Anton Souslov, and Kai Sun.
\newblock Phonons and elasticity in critically coordinated lattices.
\newblock {\em Reports on Progress in Physics}, 78(7):073901, 2015.

\bibitem{kane2014topological}
CL~Kane and TC~Lubensky.
\newblock Topological boundary modes in isostatic lattices.
\newblock {\em Nature Physics}, 10(1):39--45, 2014.

\bibitem{stenull2016topological}
Olaf Stenull, CL~Kane, and TC~Lubensky.
\newblock Topological phonons and weyl lines in three dimensions.
\newblock {\em Physical Review Letters}, 117(6):068001, 2016.

\bibitem{bilal2017intrinsically}
Osama~R Bilal, Roman S{\"u}sstrunk, Chiara Daraio, and Sebastian~D Huber.
\newblock Intrinsically polar elastic metamaterials.
\newblock {\em Advanced Materials}, 29(26):1700540, 2017.

\bibitem{paulose2015selective}
Jayson Paulose, Anne~S Meeussen, and Vincenzo Vitelli.
\newblock Selective buckling via states of self-stress in topological
  metamaterials.
\newblock {\em Proceedings of the National Academy of Sciences},
  112(25):7639--7644, 2015.

\bibitem{kadic20193d}
Muamer Kadic, Graeme~W Milton, Martin van Hecke, and Martin Wegener.
\newblock 3d metamaterials.
\newblock {\em Nature Reviews Physics}, 1(3):198--210, 2019.

\bibitem{bilal2021experimental}
Osama~R Bilal, Chern~Hwee Yee, Jan Rys, Christian Schumacher, and Chiara
  Daraio.
\newblock Experimental realization of phonon demultiplexing in
  three-dimensions.
\newblock {\em Applied Physics Letters}, 118(9):091901, 2021.

\bibitem{bertoldi2017flexible}
Katia Bertoldi, Vincenzo Vitelli, Johan Christensen, and Martin van Hecke.
\newblock Flexible mechanical metamaterials.
\newblock {\em Nature Reviews Materials}, 2(11):1--11, 2017.

\bibitem{burns1987negative}
Stephen Burns.
\newblock Negative poisson's ratio materials.
\newblock {\em Science}, 238(4826):551--551, 1987.

\bibitem{bertoldi2010negative}
Katia Bertoldi, Pedro~M Reis, Stephen Willshaw, and Tom Mullin.
\newblock Negative poisson's ratio behavior induced by an elastic instability.
\newblock {\em Advanced Materials}, 22(3):361--366, 2010.

\bibitem{yu2018mechanical}
Xianglong Yu, Ji~Zhou, Haiyi Liang, Zhengyi Jiang, and Lingling Wu.
\newblock Mechanical metamaterials associated with stiffness, rigidity and
  compressibility: A brief review.
\newblock {\em Progress in Materials Science}, 94:114--173, 2018.

\bibitem{ren2018auxetic}
Xin Ren, Raj Das, Phuong Tran, Tuan~Duc Ngo, and Yi~Min Xie.
\newblock Auxetic metamaterials and structures: a review.
\newblock {\em Smart Materials and Structures}, 27(2):023001, 2018.

\bibitem{florijn2016programmable}
Bastiaan Florijn, Corentin Coulais, and Martin van Hecke.
\newblock Programmable mechanical metamaterials: the role of geometry.
\newblock {\em Soft Matter}, 12(42):8736--8743, 2016.

\bibitem{overvelde2016three}
Johannes~TB Overvelde, Twan~A De~Jong, Yanina Shevchenko, Sergio~A Becerra,
  George~M Whitesides, James~C Weaver, Chuck Hoberman, and Katia Bertoldi.
\newblock A three-dimensional actuated origami-inspired transformable
  metamaterial with multiple degrees of freedom.
\newblock {\em Nature Communications}, 7(1):1--8, 2016.

\bibitem{lei20193d}
Ming Lei, Wei Hong, Zeang Zhao, Craig Hamel, Mingji Chen, Haibao Lu, and
  H~Jerry Qi.
\newblock 3d printing of auxetic metamaterials with digitally reprogrammable
  shape.
\newblock {\em ACS Applied Materials \& Interfaces}, 11(25):22768--22776, 2019.

\bibitem{dalela2021review}
Srajan Dalela, PS~Balaji, and DP~Jena.
\newblock A review on application of mechanical metamaterials for vibration
  control.
\newblock {\em Mechanics of Advanced Materials and Structures}, pages 1--26,
  2021.

\bibitem{cummer2016controlling}
Steven~A Cummer, Johan Christensen, and Andrea Al{\`u}.
\newblock Controlling sound with acoustic metamaterials.
\newblock {\em Nature Reviews Materials}, 1(3):1--13, 2016.

\bibitem{wu2021brief}
Lingling Wu, Yong Wang, Kuochih Chuang, Fugen Wu, Qianxuan Wang, Weiqi Lin, and
  Hanqing Jiang.
\newblock A brief review of dynamic mechanical metamaterials for mechanical
  energy manipulation.
\newblock {\em Materials Today}, 44:168--193, 2021.

\bibitem{pishvar2020foundations}
Maya Pishvar and Ryan~L Harne.
\newblock Foundations for soft, smart matter by active mechanical
  metamaterials.
\newblock {\em Advanced Science}, 7(18):2001384, 2020.

\bibitem{animate2021}
Perspective: Animate materials.
\newblock \url{www.royalsociety.org/animate-materials}, 2021.

\bibitem{susstrunk2016classification}
Roman S{\"u}sstrunk and Sebastian~D Huber.
\newblock Classification of topological phonons in linear mechanical
  metamaterials.
\newblock {\em Proceedings of the National Academy of Sciences},
  113(33):E4767--E4775, 2016.

\bibitem{shankar2022topological}
Suraj Shankar, Anton Souslov, Mark~J Bowick, M~Cristina Marchetti, and Vincenzo
  Vitelli.
\newblock Topological active matter.
\newblock {\em Nature Reviews Physics}, 4(6):380--398, 2022.

\bibitem{rocklin2016mechanical}
D~Zeb Rocklin, Bryan Gin-ge Chen, Martin Falk, Vincenzo Vitelli, and
  TC~Lubensky.
\newblock Mechanical weyl modes in topological maxwell lattices.
\newblock {\em Physical Review Letters}, 116(13):135503, 2016.

\bibitem{yang2015topological}
Zhaoju Yang, Fei Gao, Xihang Shi, Xiao Lin, Zhen Gao, Yidong Chong, and Baile
  Zhang.
\newblock Topological acoustics.
\newblock {\em Physical Review Letters}, 114(11):114301, 2015.

\bibitem{khanikaev2015topologically}
Alexander~B Khanikaev, Romain Fleury, S~Hossein Mousavi, and Andrea Alu.
\newblock Topologically robust sound propagation in an angular-momentum-biased
  graphene-like resonator lattice.
\newblock {\em Nature Communications}, 6(1):1--7, 2015.

\bibitem{mousavi2015topologically}
S~Hossein Mousavi, Alexander~B Khanikaev, and Zheng Wang.
\newblock Topologically protected elastic waves in phononic metamaterials.
\newblock {\em Nature Communications}, 6(1):1--7, 2015.

\bibitem{nash2015topological}
Lisa~M Nash, Dustin Kleckner, Alismari Read, Vincenzo Vitelli, Ari~M Turner,
  and William~TM Irvine.
\newblock Topological mechanics of gyroscopic metamaterials.
\newblock {\em Proceedings of the National Academy of Sciences},
  112(47):14495--14500, 2015.

\bibitem{ghatak2020observation}
Ananya Ghatak, Martin Brandenbourger, Jasper van Wezel, and Corentin Coulais.
\newblock Observation of non-hermitian topology and its bulk--edge
  correspondence in an active mechanical metamaterial.
\newblock {\em Proceedings of the National Academy of Sciences},
  117(47):29561--29568, 2020.

\bibitem{coulais2021topology}
Corentin Coulais, Romain Fleury, and Jasper van Wezel.
\newblock Topology and broken hermiticity.
\newblock {\em Nature Physics}, 17(1):9--13, 2021.

\bibitem{zhang2018fracturing}
Leyou Zhang and Xiaoming Mao.
\newblock Fracturing of topological maxwell lattices.
\newblock {\em New Journal of Physics}, 20(6):063034, 2018.

\bibitem{chen2014nonlinear}
Bryan Gin-ge Chen, Nitin Upadhyaya, and Vincenzo Vitelli.
\newblock Nonlinear conduction via solitons in a topological mechanical
  insulator.
\newblock {\em Proceedings of the National Academy of Sciences},
  111(36):13004--13009, 2014.

\bibitem{chen2016topological}
Bryan Gin-ge Chen, Bin Liu, Arthur~A Evans, Jayson Paulose, Itai Cohen,
  Vincenzo Vitelli, and CD~Santangelo.
\newblock Topological mechanics of origami and kirigami.
\newblock {\em Physical Review Letters}, 116(13):135501, 2016.

\bibitem{maxwell1864calculation}
J~Clerk Maxwell.
\newblock L. on the calculation of the equilibrium and stiffness of frames.
\newblock {\em The London, Edinburgh, and Dublin Philosophical Magazine and
  Journal of Science}, 27(182):294--299, 1864.

\bibitem{calladine1978buckminster}
Christopher~R Calladine.
\newblock Buckminster fuller's “tensegrity” structures and clerk maxwell's
  rules for the construction of stiff frames.
\newblock {\em International Journal of Solids and Structures}, 14(2):161--172,
  1978.

\end{thebibliography}
\end{document}